\documentclass[sigconf]{acmart}

\setcopyright{acmcopyright}
\copyrightyear{2024}
\acmYear{2024}
\acmDOI{XXXXXXX.XXXXXXX}

\acmConference[ICSE-SEIS '24]{International Conference on Software Engineering: Software Engineering in Society}{April 12--21,
  2024}{Lisbon, Portugal}
\acmPrice{15.00}
\acmISBN{978-1-4503-XXXX-X/18/06}

\usepackage[utf8]{inputenc}

\PassOptionsToPackage{hyphens}{url}\usepackage{hyperref}

\usepackage[htt]{hyphenat}

\usepackage{enumitem}
\newlist{rqs}{enumerate}{1}
\setlist[rqs]{label*=\textbf{RQ\arabic*~}}

\usepackage{csquotes}

\usepackage{booktabs,tabularx,makecell,ragged2e}

\usepackage[super]{nth}

\usepackage{graphicx}
\graphicspath{{images/}}

\usepackage{tcolorbox}
\definecolor{amber}{rgb}{1.0, 0.75, 0.0}
\newtcolorbox{mybox}{colback=gray!10,colframe=gray}

\usepackage{enumitem}

\usepackage{dblfloatfix}

\usepackage[scaled=.8]{beramono}

\begin{document}

\definecolor{colorA}{RGB}{114,141,196}

\settopmatter{printfolios=true}
\setlength{\footskip}{1.5cm}

\title[How Software Engineering Research Is Discussed on LinkedIn]{Beyond Self-Promotion:\\ How Software Engineering Research Is Discussed on LinkedIn}

\author{Marvin Wyrich}
\orcid{0000-0001-8506-3294}
\affiliation{
  \institution{Saarland University}
  \city{Saarbrücken}
  \country{Germany}
}
\email{wyrich@cs.uni-saarland.de}

\author{Justus Bogner}
\orcid{0000-0001-5788-0991}
\affiliation{
  \institution{Vrije Universiteit Amsterdam}
  \city{Amsterdam}
  \country{The Netherlands}
}
\email{j.bogner@vu.nl}

\begin{abstract}
  LinkedIn is the largest professional network in the world. As such, it can serve to build bridges between practitioners, whose daily work is software engineering (SE), and researchers, who work to advance the field of software engineering. We know that such a metaphorical bridge exists: SE research findings are sometimes shared on LinkedIn and commented on by software practitioners. Yet, we do not know what state the bridge is in. Therefore, we quantitatively and qualitatively investigate how SE practitioners and researchers approach each other via public LinkedIn discussions and what both sides can contribute to effective science communication. We found that a considerable proportion of LinkedIn posts on SE research are written by people who are not the paper authors (39\%). Further, 71\% of all comments in our dataset are from people in the industry, but only every second post receives at least one comment at all. Based on our findings, we formulate concrete advice for researchers and practitioners to make sharing new research findings on LinkedIn more fruitful.
\end{abstract}

\ccsdesc[500]{Human-centered computing~Social content sharing}

\keywords{Science Communication, LinkedIn, Software Engineering, Outreach, Knowledge Sharing, Research Impact, Community Engagement}

\begin{teaserfigure}
  \includegraphics[width=\textwidth]{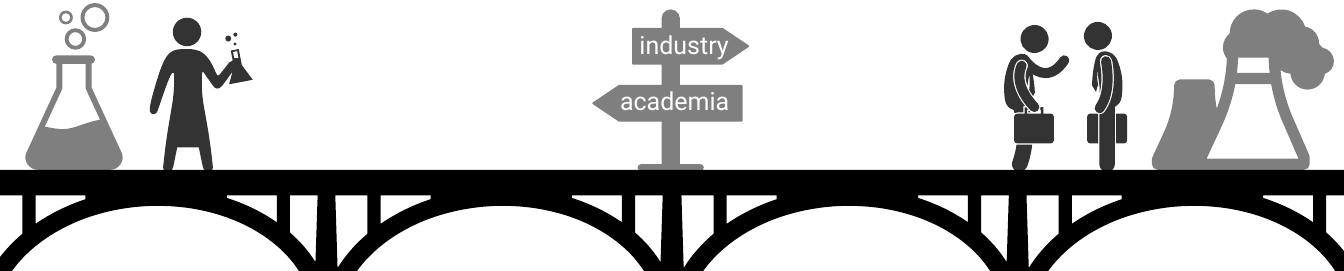}
  \caption{Metaphorical bridge on which researchers and practitioners can potentially meet and interact}
  \Description{We see a bridge that connects industry and academia. The bridge is symbolic of science communication on LinkedIn, through which researchers and practitioners can exchange knowledge. On the bridge, we see representatives of both sides, each keeping to themselves. In this paper, we investigate whether this adequately describes the situation or whether there is more exchange than depicted in this picture.}
  \label{fig:teaser}
\end{teaserfigure}


\maketitle

\section*{Lay Abstract}

LinkedIn, the world's largest professional network, isn't just for job hunting---it can also be a place where researchers and practitioners talk about scientific findings. We have looked at such interactions for people with an academic or industry software engineering (SE) background. Surprisingly, 39\% of LinkedIn posts discussing SE research aren't from the researchers themselves. And a whopping 71\% of comments on the posts come from industry professionals. Our study offers actionable insights for researchers and practitioners to further enhance LinkedIn discussions for a stronger bridge between creators and innovators in software engineering.

\section{Introduction}

Research serves society in various meaningful ways. In the context of software engineering (SE), research provides empirical insights that can improve collaboration among software professionals, or practical prototypes for tools designed to simplify everyday development.
However, research can only have an impact if it reaches its target audience. To achieve this, it is essential for researchers to make research findings accessible via various communication channels~\cite{Nisbet:2009:SciCom,Wyrich:2023:TeachingSciCom}.

One such channel is LinkedIn, \enquote{the world's largest professional network on the internet}~\cite{WhatIsLinkedIn}. LinkedIn is therefore a potentially interesting channel for science communication, as both SE researchers and SE practitioners can be found here. The two sides can thus potentially discuss relevant research findings. In the best case, this is a win-win situation: SE practitioners gain access to useful knowledge that they can integrate into their daily work. Meanwhile, SE researchers reach the audience they intend to support with their research, while receiving feedback on their work, as well as potential collaboration partners for future research projects.

At least in theory, this is how we can imagine the situation. However, we do not yet know how frequent and well such exchanges take place on LinkedIn. By analyzing this, we could first learn how practitioners respond to shared SE research, and second, make recommendations on how SE science communication could be improved by both sides. Therefore, we will investigate the following three research questions:

\begin{rqs}[leftmargin=*]
    \item How is SE research shared on LinkedIn?
    \item How do people respond to shared SE research on LinkedIn?
    \item What do positive examples of SE science communication have in common?
\end{rqs}

Following \citet{Burns:2003:SciComDefinition}, we define science communication (SciCom) \enquote{as the use of appropriate skills, media, activities, and dialogue to produce one or more of the following personal responses to science: Awareness, Enjoyment, Interest, Opinion-forming, and Understanding}~\cite{Burns:2003:SciComDefinition}.
It is easy to find examples of SE researchers using LinkedIn as a medium to at least create \emph{awareness} for their research papers and artifacts. However, so far, we can only guess as to how well this works.

Anecdotally, SE researchers report that science communication on LinkedIn can be quite successful. For example, Marcos Kalinowski wrote on social media: 
\enquote{Recently I shared the result of a PhD thesis on LinkedIn and it reached 4,000+ reactions and 270,000+ impressions}~\cite{MarcosTweet}.
He further emphasizes an aspect that certainly creates a mix of delight and optimism among a part of the SE research community: \enquote{95\% of my network is from industry. We are well equipped to burst the academic bubble!}~\cite{MarcosTweet}.

To find out if this was just a one-hit wonder among a few success stories, we systematically searched and analyzed existing LinkedIn posts in which SE research papers and findings have been shared.
In Section~\ref{sec:methodology}, we describe our approach to answering the three research questions. We then present our findings in Section~\ref{sec:results}, which we enrich with insights from related work in Section~\ref{sec:related_work}. In Section~\ref{sec:discussion}, we provide concrete recommendations on how SE researchers and SE practitioners can benefit more effectively from science communication on LinkedIn in the future. Finally, Section~\ref{sec:conclusion} concludes the paper.

\section{Methodology}
\label{sec:methodology}

Our methodology for answering the three research questions can be divided into finding relevant LinkedIn posts and analyzing these posts quantitatively and qualitatively.
We describe our approach in detail in the following subsections.
For transparency and reproducibility, we share our study artifacts online.\footnote{\url{https://doi.org/10.5281/zenodo.8410796}}

\subsection{Research Context}

LinkedIn\footnote{\url{https://www.linkedin.com}} is a social network that is primarily intended for networking in a professional work context.
LinkedIn became part of Microsoft in 2016.
The authors of this paper have LinkedIn accounts themselves, but are not otherwise affiliated with LinkedIn or Microsoft.

Anyone can create an account and set up a profile.
Thus, both SE researchers and practitioners can be found on LinkedIn, which holds the potential to engage in conversation with each other.
LinkedIn users can create content in the form of \textit{posts}.
While posts are public by default, they are mostly seen by the poster's network, as posts appear in the feeds of all connections that a poster has.
Other users can respond in several ways to a post. 
The most common, least-effort option is using emoji-based \textit{reactions} such as \texttt{like}, \texttt{celebrate}, \texttt{support}, \texttt{love}, \texttt{insightful}, or \texttt{funny}.
Another low-effort response is \textit{reposting}, i.e., directly sharing the post with your own connections, which typically indicates that you find the content relevant for other people in your network.
The last type of response, and also the one requiring the most effort, is \textit{commenting}, i.e., adding a textual response below the post.
In the context of this study, we analyze posts about specific SE research papers and how people respond to them.

\subsection{Sampling and Filtering}

The first step was to find LinkedIn posts that mention a concrete SE research paper. A mention is considered sufficient if within the LinkedIn post a) either the title of the paper and a link to the paper or artifact repository is shared, or b) at least one sentence about the content of the paper and a link is shared, or c) at least two sentences about the content of the paper are shared. Posts written by abstract-posting bots were excluded. All three options set the bar for science communication rather low. However, we deliberately refrained from qualitative filtering at this point to capture as accurate a picture as possible of the status quo of how SE research is currently shared on LinkedIn.

To assess whether the paper was about SE research, we restricted ourselves to publications from the two largest and most prestigious SE conferences: the International Conference on Software Engineering (ICSE)\footnote{\url{https://conf.researchr.org/series/icse}} and the Joint European Software Engineering Conference and Symposium on the Foundations of Software Engineering (FSE)\footnote{\url{https://conf.researchr.org/series/fse}}. 
Our process for finding relevant posts was as follows:

\begin{enumerate}
    \item Obtain a list of ICSE publications from IEEEXplore
    \begin{itemize}
        \item Tracks: technical, SEIS, SEET, SEIP
        \item Years: 2023, 2022, 2021, 2020, 2019, 2018
    \end{itemize}
    \item Obtain a list of FSE publications from ACM Digital Library
    \begin{itemize}
        \item All papers labeled as \enquote{conference paper} in the \mbox{EndNote} export, i.e., no \enquote{software} entries
        \item Years: 2022, 2021, 2020, 2019, 2018 (The 2023 proceedings were not available at the time of data collection and were therefore not included in this step)
    \end{itemize}
    \item For each paper title, perform a search on LinkedIn and save the URL of each result.
    \item Perform a keyword search on Google to find additional relevant LinkedIn posts. We used six different queries that were restricted to the LinkedIn domain, searching for conference names and their acronyms (see Table~\ref{tab:google-search-strings}). The final queries emerged from iterative tests with different keywords and combinations. Together, these search strings produced 830 results. In this step, it would have been possible to include LinkedIn posts on FSE 2023 papers, but it did not happen.
\end{enumerate}

Two authors then independently went through all the search results of this process, decided on inclusion, and finally discussed differences until consensus was reached in a joint meeting.
At this point, we had a list of relevant LinkedIn posts about concrete ICSE and FSE papers. We executed our search for posts on August \nth{16}, 2023. 

\begin{table}[ht]
    \centering
    \caption{Google search strings to identify LinkedIn posts}
    \begin{tabularx}{\columnwidth}{Xr}
        \toprule
        Search term & \# of results\\
        \midrule
        (1) \texttt{site:linkedin.com "icse paper"} & 35\\
        (2) \texttt{site:linkedin.com/posts "at icse"} & 480\\
        (3) \texttt{site:linkedin.com/posts "at the international conference on software engineering"} & 267\\
        (4) \texttt{site:linkedin.com "esec/fse paper"} & 0\\
        (5) \texttt{site:linkedin.com/posts "at esec/fse"} & 38\\
        (6) \texttt{site:linkedin.com/posts "european software engineering conference"} & 10\\
        \bottomrule
    \end{tabularx}
    \label{tab:google-search-strings}
\end{table}

\subsection{Data Collection and Analysis}
For each included post, we collected a total of 19 attributes related to the author of the LinkedIn post (\textit{poster}), the LinkedIn post itself, and the comments on each LinkedIn post.
Several easy-to-collect attributes were extracted by a Python script, e.g., the post date, the poster name and description, the \# of characters of the post, or the \# of reactions, reposts, and comments.
For more complex attributes, the extraction could not be automated with reasonable effort or accuracy and needed to be performed manually, e.g., for the poster affiliation (industry or academia), if the post contained the paper link, the post intention, or the comment intentions.
These extractions were performed according to a four-eyes principle: two authors independently extracted and subsequently compared the attributes.
In several meetings, the results of the extraction were then discussed to reach consensus for further analysis steps.

The collected data was analyzed with a mix of quantitative and qualitative methods.
Three attributes required qualitative coding: the intention of the LinkedIn post, the description of the images included in the post, and the intention of the respective comments on a LinkedIn post.
We followed thematic synthesis guidelines~\cite{Cruzes2011} to iteratively form and refine our categories.
The resulting categories are mentioned in the results section for the respective analysis and are further listed together with their definitions in the supplemental materials.
For the numeric attributes, we used basic descriptive statistics such as mean and median to analyze the posts (RQ1) and the responses to them (RQ2).
We also created various diagrams to visualize distributions, such as histograms or stacked bar charts.
In many cases, we also revisited the LinkedIn posts directly to understand their structure, especially for RQ3.
While we compared distributions between different attributes, e.g., posts from academia vs. posts from industry, no inferential statistics or hypothesis testing was used.
The primary goal of our analysis was to describe the current state of SE SciCom on LinkedIn, and not to analyze relationships or predictors.

\subsection{Limitations}
Our study design has a few limitations that are relevant for the interpretation of the results and that we would therefore like to point out in advance.
Regarding \textit{sampling}, LinkedIn makes it unfortunately very difficult to find relevant posts.
Its API does not provide search functionality for posts, and the web interface for searching posts is also very limited.
The provided search results seem to go back only a fixed timeframe, which made identifying older posts from several years ago more difficult.
Moreover, URLs in posts are shortened, which makes searching for concrete paper URLs impossible.
To partially mitigate this, we additionally used Google search.
Nonetheless, we cannot make reliable statements about the percentage of SE papers that are shared on LinkedIn, which would require a different sampling approach.
We also focused exclusively on the two most prestigious SE conferences (ICSE and FSE), which could have an effect on how SciCom manifests for the papers.
There are many more SE journals and conferences which we did not analyze, and we have to be careful with generalizations.
Still, we believe that SciCom for the top venues in a field should be held to a high level of expectations.

Regarding \textit{data collection}, a small percentage of posts were edited and therefore could have added the paper link later on.
We selected and extracted posts based on their content at the time of extraction, not their unedited content.
As a result, our dataset could contain a few posts that would otherwise have been excluded for not containing a paper link.
Similarly, when categorizing posts and comments, we took the current affiliation of the author, and did not check if this affiliation was also the one active when the comment was made. In the vast majority of cases, the affiliation or at least whether it was in industry or academia should be the same, but it is possible that a small percentage of posts and comments were categorized incorrectly because of this.
For the automatic extraction of attributes, we carefully developed our script and double-checked the results for a small sample.
While we did not find any errors, the UI-based script might have extracted wrong numbers for unidentified edge cases.
We believe the impact of both these threats to be negligible.

Lastly, LinkedIn posts are shared within the network of the poster and the networks of those reposting.
The content and quality of a post are therefore not the only influencing factors for, e.g., the number of comments on posts or the contents of the comments.
We therefore have to be careful to generalize conclusions regarding effective SciCom on LinkedIn based on our data, as we do not include, e.g., the number of followers or connections per poster.

\begin{figure*}[htbp]
    \centering
    \includegraphics[width=\textwidth]{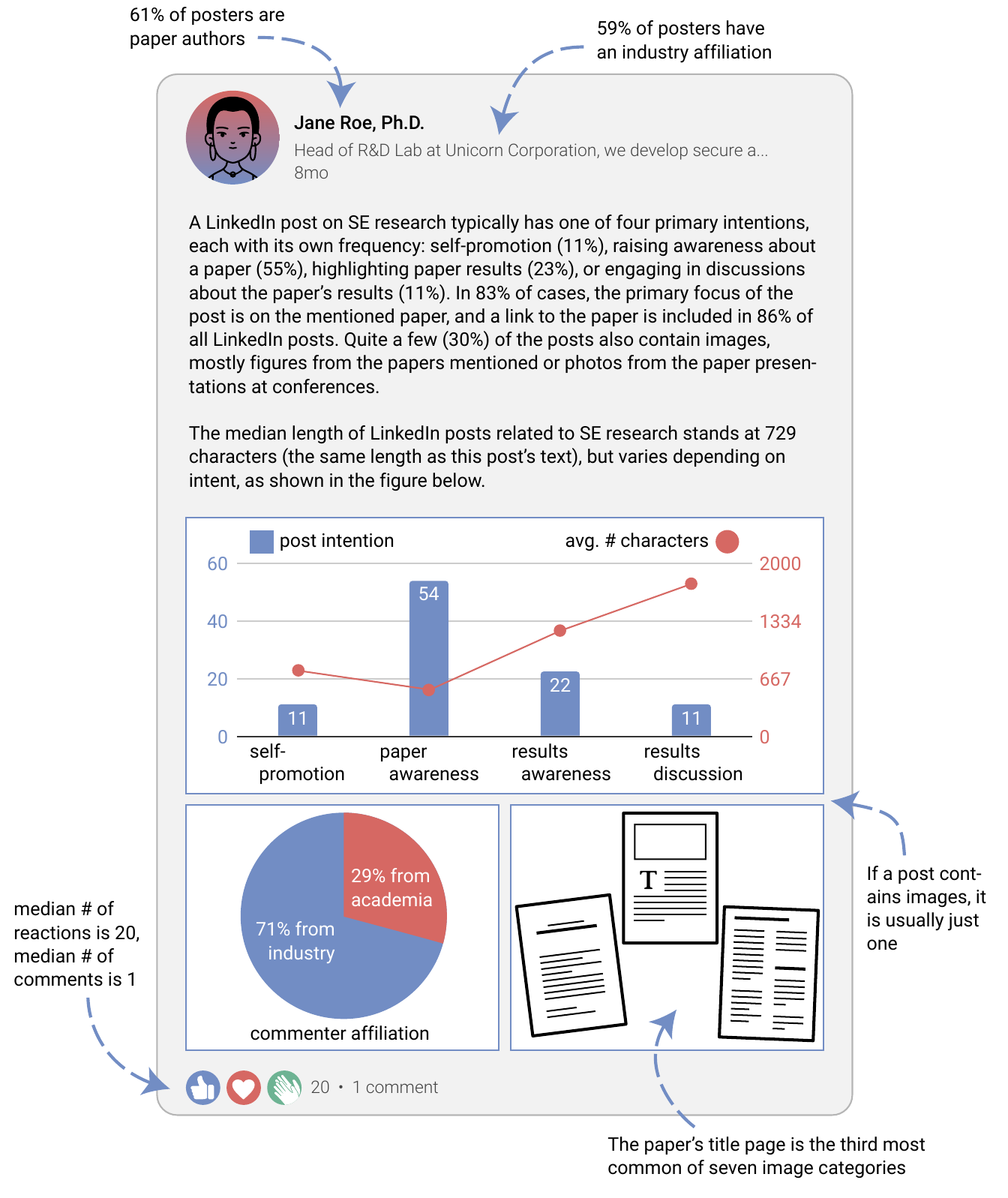}
    \caption{Mockup of a LinkedIn post with some key results for our research questions}
    \label{fig:rq1_results}
\end{figure*}

\section{Results}
\label{sec:results}

Our sample includes a total of 98 LinkedIn posts. These posts were written by 86 unique posters, of which 35 (40.7\%) had an academic affiliation and 57 (59.3\%) had an industry affiliation. Almost all posts in our sample were written in English (96/98), with one exception in Portuguese and one in Dutch.
The LinkedIn posts mentioned 79 different software engineering papers, covering a broad thematic spectrum ranging from technical innovation to human-centered engineering. The first author of these papers was mostly affiliated with academia (75.9\%). Interestingly, only 61.2\% of LinkedIn posts were created by a paper author.

In the following three subsections, we investigate how the 98 LinkedIn posts on SE research can be characterized (RQ1), how people responded to these posts (RQ2), and which examples of LinkedIn posts stood out in their science communication (RQ3).

\subsection{Characteristics of LinkedIn Posts (RQ1)}

To provide the reader with a better sense of what a LinkedIn post generally looks like, we have mocked up a LinkedIn post in Fig.~\ref{fig:rq1_results}, which we also use to directly present some of our key results.

With 850 characters ($SD=646$), an average LinkedIn post on SE research comprises about one to two paragraphs. In terms of content, 81 (83\%) of the posts focus on a scientific paper. In the other 17 posts, a paper published at ICSE or ESEC was also mentioned, but the focus of the post was on other concerns, such as receiving an award for the paper or describing the experience of a conference visit to present the paper.
In our analysis, we categorized all 98 posts into one of four intentions: self-promotion, paper awareness, results awareness, and results discussion.

\begin{enumerate}
    \item \texttt{self-promotion:} the primary goal of the post is to promote one's own achievements or organization, e.g., getting accepted at a prestigious conference or winning an award
    \item \texttt{paper awareness:} the primary goal of the post is to raise awareness about the existence of a paper and its high-level topic, e.g., to encourage others to read and cite it
    \item \texttt{results awareness:} the primary goal of the post is to raise awareness about more concrete results of the paper, e.g., important findings / recommendations or developed tools
    \item \texttt{results discussion:} the primary goal of the post is to present the paper results in more details, ideally in an easily digestible manner, to discuss their implications or interpretations, often with the intention to start a discussion about the topic via comments
\end{enumerate}

It is important to note that these four categories should be understood in a hierarchical manner, wherein, for instance, \texttt{results awareness} encompasses \texttt{paper awareness} and \texttt{self-promotion}.
In Fig.~\ref{fig:rq1_results}, we present a diagram showing the frequency of the four post intention categories, and we see that LinkedIn post authors most often draw attention to a paper (54/98), and occasionally describe the results of a paper (22/98). Less frequently, the results of a paper are discussed (11/98), for example in the form of highlighting practical takeaways or by enriching the results with the experience of the post author.
Similarly, it is rare for the main intention of a LinkedIn post to focus solely on the poster's achievement (\texttt{self-promotion}), for example, when the acceptance of the paper rather than its findings is the main concern (11/98).

In addition to textual content, 29 (30\%) of the LinkedIn posts also included at least one image. We divided the total of 43 images into seven categories: \texttt{paper figure} (13), \texttt{presentation photo} (11), \texttt{paper title page} (7), \texttt{visual hook} (5), \texttt{event photo} (3), \texttt{event announcement} (2), \texttt{non-paper infographic} (2). Thus, in summary, in the cases where an image was shared, it was mostly either taken from the paper being discussed, it was a screenshot of the first page, or it was an image of the paper author's conference presentation. In five cases, the poster created an image that was intentionally used to draw attention to the LinkedIn post (\texttt{visual hook}), and in two cases, an informative graphic was created that did not appear in the paper.

Descriptions of the intention and image categories, as well as the categorization of each of the 98 LinkedIn posts, are provided in the supplemental materials.

\subsection{Responses to LinkedIn Posts (RQ2)}

\begin{figure*}[ht]
  \centering
  \begin{minipage}[b]{0.47\textwidth}
    \includegraphics[width=\columnwidth]{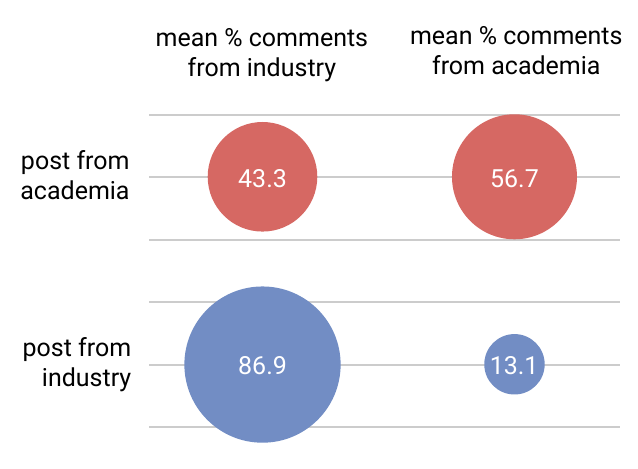}
    \caption{Comments on posts from academia vs. industry}
    \label{fig:comments-ind-vs-aca}
  \end{minipage}
  \hfill
  \begin{minipage}[b]{0.47\textwidth}
    \includegraphics[width=\columnwidth]{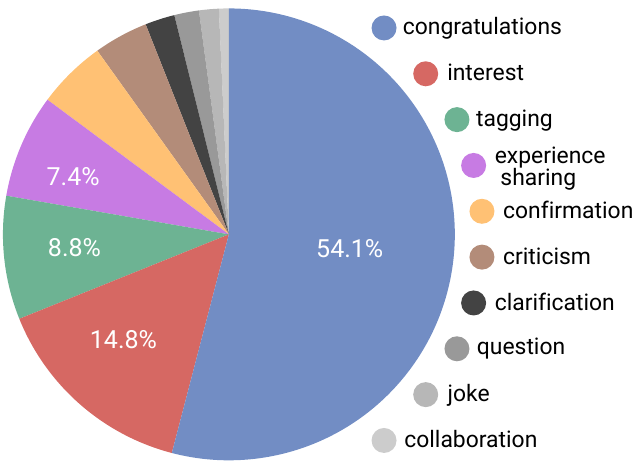}
    \caption{Distribution of intentions in non-author comments}
    \label{fig:comment-intentions}
  \end{minipage}
\end{figure*}

\begin{figure*}[b]
    \centering
    \includegraphics[width=\textwidth]{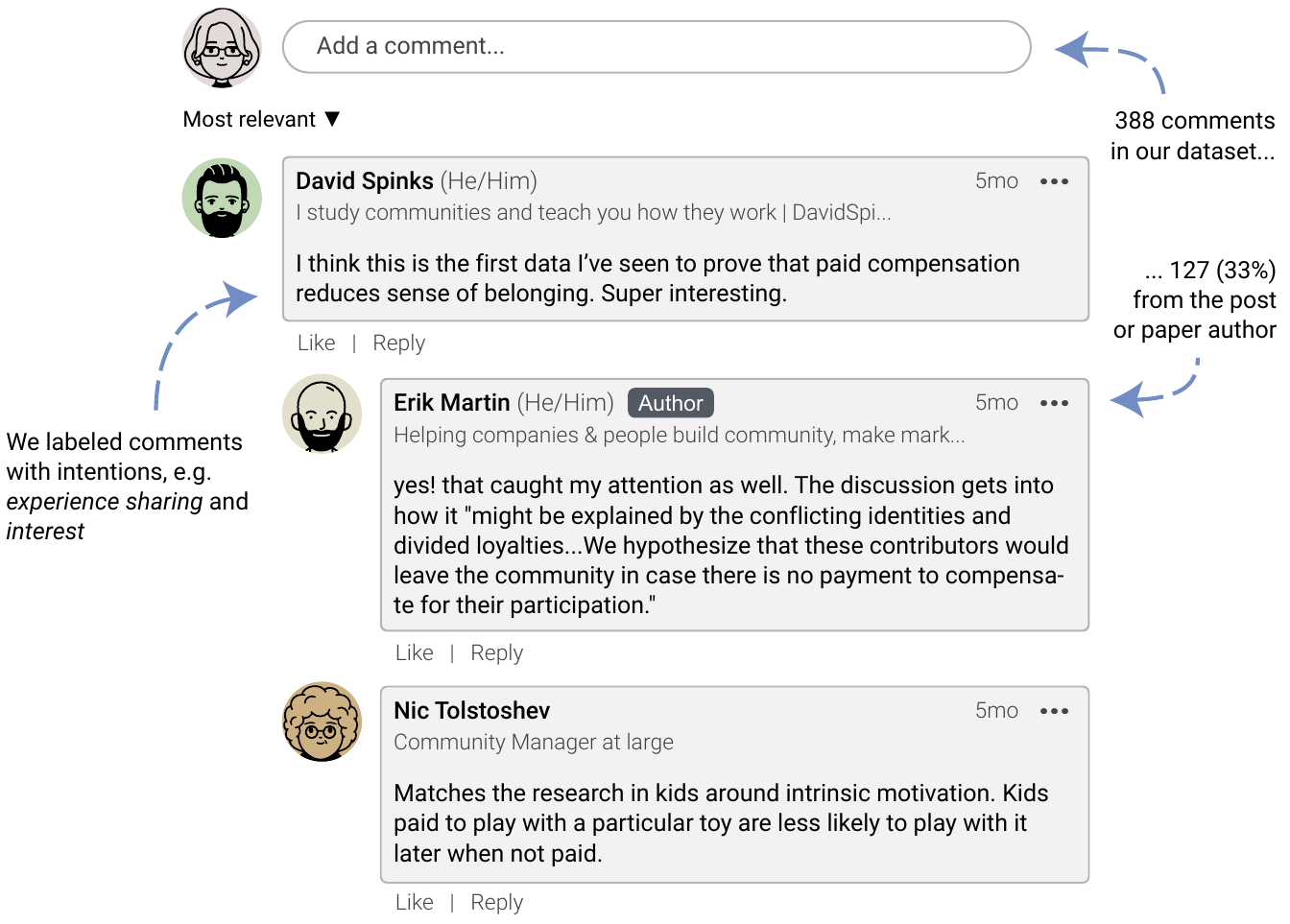}
    \caption{Example of a comment thread from LinkedIn post with \href{https://www.linkedin.com/posts/erikmartin_community-communitymanagement-devrel-activity-7050192077636161536-XKQq}{\color{blue}ID 24} on a paper about belonging in open source}
    \label{fig:comment_thread_example}
\end{figure*}

On LinkedIn, people have several options to respond to posts.
We analyzed \textit{reactions \& reposts} plus \textit{comments} and their \textit{intentions}.

\subsubsection{Reactions \& Reposts}
Each post in our sample received at least one reaction, with a median of 20.5 and a mean of 47.1 reactions.
One extremely popular outlier post (ID 2) received a total of 1,185 reactions, which explains the difference between mean and median.
In general, this type of response was fairly frequent in our dataset, with only 14 posts receiving 5 or fewer reactions.
However, \textit{reposting} was considerably less common in our sample.
From our 98 posts, 53 did not receive a single repost (54\%), 28 were reposted once (29\%), and only 17 received 2 or more reposts (17\%).
On average, each post received 2.1 reposts, with a median of 0.
The mean was again inflated by the popular outlier (ID 2) that received 101 reposts.
All in all, people made much less use of reposts than reactions.

We also compared these responses for posts from industry vs. academia and posts from paper authors vs. non-authors.
The outlier post (ID 2) was removed for this analysis, as it had a zscore > 3 for both reactions and reposts.
All in all, posts from academia in our sample attracted more reactions than posts from industry (mean: 47.0 vs. 26.9, median: 37 vs. 14.5) and slightly more reposts (mean: 1.3 vs. 0.9, median: 1 vs. 0).
Similarly, posts from paper authors also attracted more reactions (mean: 43.3 vs. 22.5, median: 37 vs. 11) than non-author posts.
However, the same was not true for reposts, as non-author posts were reposted slightly more often (mean: 1.4 vs. 0.9, median: 1 vs. 0).

\begin{figure*}[b!]
    \centering
    \textcolor{white}{\rule[1ex]{\textwidth}{0.5pt}}
    \includegraphics[width=\textwidth]{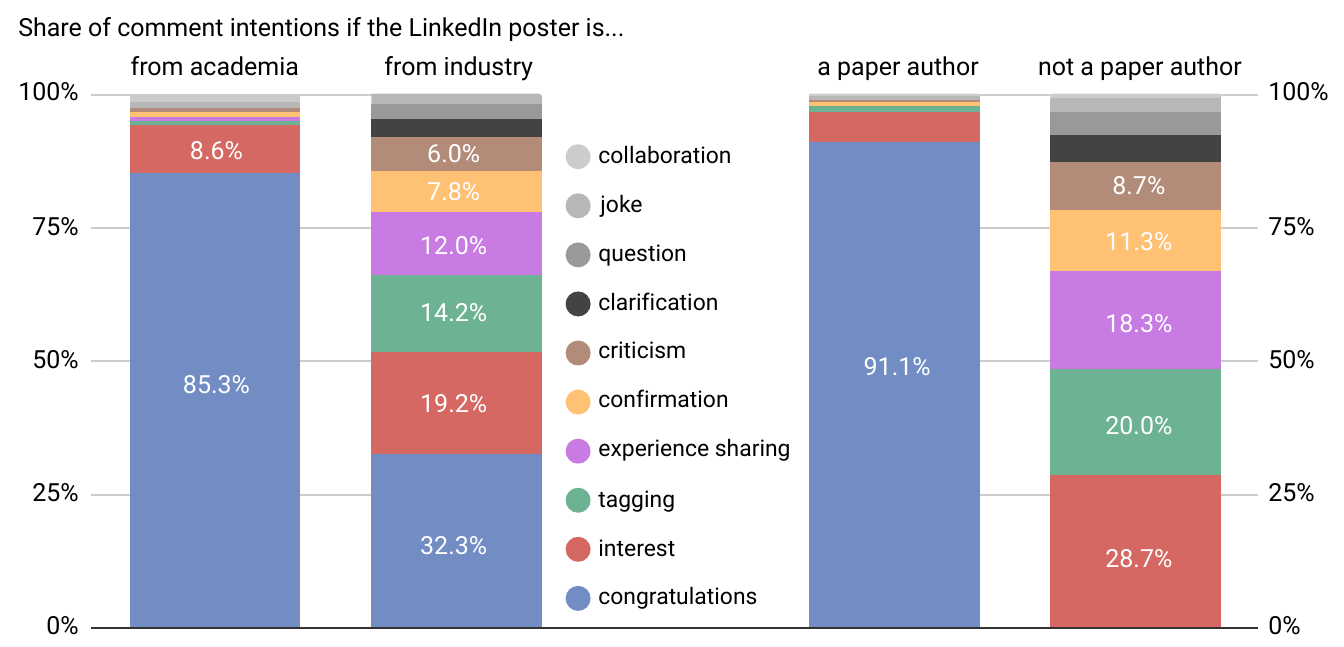}
    \caption{Comment intentions by characteristics of the LinkedIn poster}
    \label{fig:comment-intention-comparison}
\end{figure*}

\subsubsection{Comments}
Our dataset of 98 posts contains a total of 388 comments.
However, 127 comments are from the post authors or from authors of the shared paper themselves (33\%).
Since we are primarily interested in how \textit{other} people respond to the shared SE research, we excluded these comments from the analysis, i.e., our sample contains 261 comments.
All in all, comments were only really prevalent on a small subset of posts.
Nearly half of our sample received zero comments (48\%), with another 28 posts only having 1 or 2 comments (29\%).
More than 2 comments only appeared for 23 of 98 posts (23\%).
This led to a mean of 2.7 comments per post, with a median of 1.
In general, posts from academia tended to attract slightly more comments than industry posts (mean: 2.8 vs. 2.6; median: 1 vs. 0), with the same being true for posts shared by a paper author (mean: 2.8 vs. 2.5; median: 1 vs. 0).

Regarding the affiliation of the people who commented, 71\% of comments in our sample were from industry and 29\% from academia.
From a SciCom perspective, this seems great because, as a research community, we want industry to see and engage with SE research.
However, when only taking the 51 posts that received at least one comment and grouping between posts from industry and academia, we see an interesting pattern (Fig.~\ref{fig:comments-ind-vs-aca}).
Posts from academia roughly received a 40-60 split between comments from industry vs. academia, which seems fairly beneficial.
Conversely, posts from industry attracted the vast majority of comments only from industry (87\%).
In the median, zero of the existing comments on industry posts were from academia (mean: 0.7).
This seems to indicate that if industry shares academic research, other practitioners will see and comment on it much more, but the discussion appears to be more siloed, without academic involvement beyond potential paper authors.

\subsubsection{Comment Intentions}

We also analyzed the \textit{intention} of comments and derived several labels.
Each comment was labeled with at least one of the following intentions:

\begin{itemize}
    \item \texttt{collaboration}: requesting to collaborate or discussing previous collaborations
    \item \texttt{interest}: voicing interest in the topic or indicating the paper's helpfulness
    \item \texttt{confirmation}: voicing support for the paper's results
    \item \texttt{criticism}: raising criticism of the research approach or voicing skepticism about the paper's results
    \item \texttt{question}: asking a question about the topic or paper
    \item \texttt{experience sharing}: sharing experiences related to the paper topic
    \item \texttt{clarification}: answering questions or correcting other comments
    \item \texttt{tagging}: tagging people from their network to share the post
    \item \texttt{congratulations}: congratulating the paper author for the paper acceptance or for winning an award
    \item \texttt{joke}: making a funny remark or joke
\end{itemize}

\newpage
From a SciCom perspective, some of these intentions signify high-value interactions, e.g., \texttt{question}, \texttt{interest}, \texttt{experience sharing}, or \texttt{criticism}, while others represent comments of rather low potential for interactions, e.g., \texttt{congratulations} or \texttt{joke}.
Fig.~\ref{fig:comment-intentions} summarizes the distribution of comment intentions in our sample.
More than half of the 261 comments (54\%) were simple \texttt{congratulations} on getting a paper accepted or winning an award, and therefore did not add much value to the discussion of the research.
Comments with more potential for SciCom interactions signified \texttt{interest} (15\%), \texttt{tagged} other people to make them aware (9\%), or \texttt{shared experiences} related to the topic (7\%).
Comments voicing \texttt{confirmation} (5\%) or \texttt{criticism} (4\%) were rare, with comments about \texttt{clarifications} (2\%), \texttt{questions} (2\%), \texttt{jokes} (1\%), or \texttt{collaborations} (1\%) only appearing very sporadically.
Fig.~\ref{fig:comment_thread_example} shows a concrete example of a comment thread.

When analyzing on which posts low-potential comments were more likely to appear, we found substantial differences between posts from academia vs. industry and posts from paper authors vs. non-authors.
The vast majority of high-potential comments such as \texttt{interest}, \texttt{tagging}, \texttt{experience sharing}, \texttt{criticism}, etc. appeared on posts from industry.
For example, academic posts were responsible for 100 of the 153 \texttt{congratulation} comments despite only 42\% of the 98 posts being from academia.
The same pattern appeared even more strikingly for posts by non-paper authors, which generally attracted the most valuable comments from a SciCom perspective.
For example, 100\% of \texttt{congratulation} comments appeared on paper author posts, while \texttt{experience sharing} happened exclusively with non-author posts.
Fig.~\ref{fig:comment-intention-comparison} visualizes these differences.

\begin{table*}[b]
    \centering
    \caption{Positive Examples of SE SciCom on LinkedIn}
    \begin{tabular}{rllrrrrr}
        \toprule
        ID & URL & Post intention & \# of characters & \# of images & \# of reactions & \# of reposts & \# of non-author comments\\
        \midrule
        2 & \href{https://www.linkedin.com/posts/abinoda_developerexperience-softwareengineering-developerproductivity-activity-7004477107124789248--yed/}{\color{blue}linkedin.com}  & results awareness & 954 & 1 & 1,185 & 101 & 49 \\
        24 & \href{https://www.linkedin.com/posts/erikmartin_community-communitymanagement-devrel-activity-7050192077636161536-XKQq/}{\color{blue}linkedin.com} & results discussion & 2,388 & 0 & 81 & 6 & 17 \\
        40 & \href{https://www.linkedin.com/posts/jpatrickhall_pytorch-deeplearning-deeplearning-activity-7007387782037782528-a4Pp/}{\color{blue}linkedin.com} & results discussion & 2,565 & 1 & 200 & 13 & 5 \\
        42 & \href{https://www.linkedin.com/posts/junweih_carnegiemellon-ml-datateam-activity-6977636326443552768-_fEt/}{\color{blue}linkedin.com} & results discussion & 2,256 & 1 & 142 & 18 & 7 \\
        86 & \href{https://www.linkedin.com/posts/walid-maalej-b885068_new-research-paper-recently-accepted-for-activity-7054768701463515136-EW_w/}{\color{blue}linkedin.com} & results discussion & 2,891 & 1 & 38 & 0 & 2 \\
        93 & \href{https://www.linkedin.com/posts/yaroslav-golubev_research-softwareengineering-codequality-activity-7026181178470641664-wraX/}{\color{blue}linkedin.com} & results discussion & 1,275 & 3 & 10 & 1 & 0 \\
        \bottomrule
    \end{tabular}
    \label{tab:good-examples}
\end{table*}

\subsection{Positive Examples (RQ3)}
Lastly, we wanted to identify and compare positive examples of SE SciCom on LinkedIn.
To find these posts, we repeatedly sorted our sample by several attributes that could indicate the success or quality of a post: post intention, \# of characters, \# of images, \# of reactions, \# of reposts, and \# of non-author comments (minus \texttt{congratulations}).
While high values in these metrics did not guarantee a high-quality LinkedIn post, we found them to be good indicators to then manually assess and identify positive examples in our sample.
Based on this analysis, we selected six posts that are listed in Table~\ref{tab:good-examples}.

\begin{figure}[t]
    \centering
    \includegraphics[width=\columnwidth]{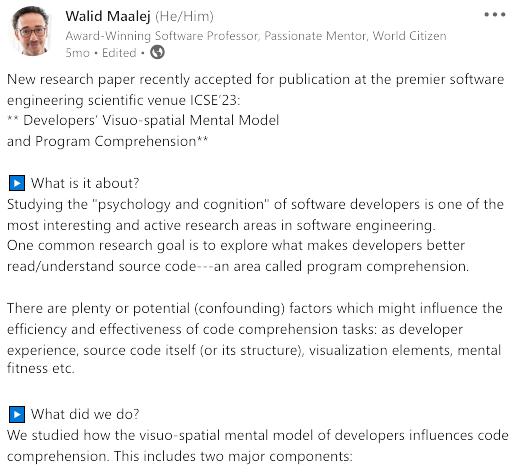}
    \caption{Positive example of SE SciCom on LinkedIn (\href{https://www.linkedin.com/posts/walid-maalej-b885068_new-research-paper-recently-accepted-for-activity-7054768701463515136-EW_w/}{\color{blue}ID 86})}
    \label{fig:post-example}
\end{figure}

All of these posts share several characteristics.
Most of them received the \textit{results discussion} label as their intent, with ID 2 being the only exception with \textit{results awareness}.
The findings or key results of the papers are not only abstractly hinted at or briefly mentioned, but presented with sufficient detail to understand their significance or relevance.
Additionally, the posters usually added their own interpretation to guide understanding and to help readers grasp the industrial relevance.
All except one post (ID 24) also included at least one image to further support their message, usually a key results figure from the paper.
However, this level of detail also led to longer posts, with (sometimes substantially) more characters than the median of 729.
To still ensure readability despite this text length, the posters made good use of bullet points, paragraphs, and headings to structure the content.
A nice example of this is provided in the post with ID 86: despite being the longest post in our sample, the text is effectively broken up into manageable sections with headings that make it clear what each section is about (see Fig.~\ref{fig:post-example}).
Structuring and supporting the text in such a way is crucial to not lose potential readers immediately due to a stereotypical \enquote{wall of text}.
Several posts in our general sample were of comparable length, but simply copied large parts of text from the paper, which is unsuitable for the LinkedIn audience.

From the six selected positive examples, only two were posted by paper authors (IDs 86 and 93) and only two from people from academia (IDs 40 and 86).
In general, posters from industry seemed more capable of presenting the papers in an engaging and easily digestible manner, which often led to numerous responses.
For example, the posts with IDs 2 and 24 managed to attract 49 and 17 comments respectively that were all more than the simple congratulations that we often saw on academic posts.
Several positive examples also reached high numbers of reactions and reposts.
However, the posts with IDs 86 and 93 performed not as well in these metrics in comparison.
This highlights the fact that you can create a high-quality LinkedIn post from a SciCom perspective, but still not reach a substantial number of interactions.
Having a large network and a paper topic that many practitioners find interesting undoubtedly plays an even larger role here.
For example, the posts with IDs 2, 24, 42 were written by authors with extensive networks of several thousands of followers.
Additionally, the high-performing posts also were about paper topics that are generally interesting for many SE practitioners, e.g., developer productivity (ID 2), sense of community in development teams (ID 24), or machine learning (IDs 40 and 42).
Conversely, the post with ID 93, despite being a good example of SE SciCom, was about academic SE education, which is not a high-interest topic for most practitioners.

\section{Related Work}
\label{sec:related_work}

While we are not aware of any study that analyzes how SE research is communicated on social media, there are a few papers related to this topic.
For example, there is a fair amount of research on how SE professionals use social media and microblogging for collaborative software development, community building, technical education, and other social interactions~\cite{Begel:2010:SocialMedia, Storey:2010:ImpactSocialMedia, Storey:2014:RevolutionSocialMedia, Barzilay:2011:LinkedInExamples, Tian:2012:Microblog, Singer:2014:Twitter, Bougie:2011:Twitter, Sharma:2015:WhatsHot}.
This is primarily about the exchange among practitioners, who are also present on LinkedIn in all their professional diversity: with different job titles, employed by different large companies, and with different programming experience~\cite{Barzilay:2011:LinkedInExamples}.
However, it is rarely mentioned in these research studies that \emph{researchers} could also make use of social media to share their research findings. After all, researchers would have nothing less than \enquote{the opportunity through social media to influence and guide the industry}~\cite{Storey:2014:RevolutionSocialMedia}.

Closer to the SE SciCom subject are \citet{Wyrich:2023:TeachingSciCom} who recently published an experience report on a science communication seminar for computer science students at their university. Each student was assigned a published SE paper and then had to adopt the perspective of a researcher who wants to design a communication strategy to effectively distribute the findings among the target audience(s). The students came up with different ideas for how SE research findings could be communicated, some of which were similar to LinkedIn posts~\cite{Wyrich:2023:TeachingSciCom}. The seminar participants were undergraduates, but we believe that such SciCom training would be particularly worthwhile for PhD students to establish themselves more effectively within and outside their research community. We will return in our discussion to how LinkedIn posts can have a potentially greater impact with just a few tweaks.

Communication between academia and industry can also take place much earlier than after the publication of a paper. 
\citet{Rico:2021:CaseInduAca} demonstrated through a case study how \enquote{conducting joint research both requires and contributes to bridging the communication gap between industry and academia (IA) in software engineering}. Although the context of our study is post-publication communication, some of the findings of \citeauthor{Rico:2021:CaseInduAca} can certainly be transferred: for example, \emph{research relevance} and \emph{practitioner's attitude towards research} can be drivers for communication~\cite{Rico:2021:CaseInduAca}. 
Regarding research relevance, we made a similar observation when searching for successful LinkedIn posts (RQ3): high-performing posts were about topics that are generally interesting for many SE practitioners.

Regarding the aspect of practitioner's attitude towards research, \citet{Jones:2019:RScience} observed that active participants in scientific discussions typically are those already invested in science. They analyzed 1,859 posts on the Reddit forum r/science and interviewed 18 community members to find out who the participants were and what drove their participation. They found that comments, in most cases, served the exchange of information. Furthermore, the comments were crucial in determining whether users engaged with the research at all---meaning that \enquote{the comments section itself becomes the primary artifact that communicates the science, instead of the linked articles}~\cite{Jones:2019:RScience}. Similar research strongly suggests that the difficulty and quality of comments to a news article affects people's interest in reading the article~\cite{Williams:2021:EffectsOfComments}.
The situation could be similar with LinkedIn comments, although we have little evidence on this in our data. While we would expect that the content of the LinkedIn post in which the article is shared to have the greater effect on engagement with the article, we leave these hypotheses up to future work.

Finally, to be able to write a LinkedIn comment, there must first be a LinkedIn post. Therefore, one final aspect we would like to address in this section is the motivation of people who share scientific content.
\citet{Jones:2019:RScience} also showed with their study on SciCom on Reddit that Computer Science is among the least frequent high-level post topics (0.8\%) and that researchers in general are anxious about participating in discussions about their work~\cite{Jones:2019:RScience}.
Our data seems to confirm a certain reluctance because we could only find LinkedIn posts for a small fraction of SE research published in our two top conferences. Furthermore, we found a surprisingly high number of LinkedIn posts coming from posters who are not paper authors (39\%).
Investigating the intentions of this particular group of posters presents an exciting research opportunity. \citet{Hu:2022:MotivationSecondary}, for example, found the following predictors for the intentions to participate in secondary science communication on social media: \enquote{emotional value, social value, altruistic value, attitude, internal perceived behavioral control and subjective norm}. Whether some of these can also be found for SE science communicators remains to be shown.
What we already know, however, is that there is currently little for SE researchers to fear when sharing their research on LinkedIn. Paper authors almost exclusively receive encouragement in the form of comments congratulating or people expressing interest. Moreover, the few critical comments in our dataset were always aimed at the (supposed) content of the research, such as doubted premises or critical questions about the methodology.
We did not find any particularly negative or \emph{hate} comments, which are often part of social media~\cite{Castano:2021:Hate} and can sometimes be attributed to people's attitude towards science~\cite{Batchelor:2023:JustClickbait}.

\section{Discussion}
\label{sec:discussion}

Science communication helps build a bridge between SE researchers and practitioners that ultimately benefits people on both sides. We argue that effective SciCom is the natural extension of open science~\cite{Mendez:2020:OpenScience}: if your research is not noticed by the target audience, then it does not help much that it is openly available. 

As we have seen, LinkedIn is a relevant communication channel to draw attention to SE research. Much of this communication already works really well, as we have witnessed some nice examples of science communication and active discussions among practitioners under the respective posts. However, we also see potential for a few improvements. In the following subsections, we will discuss the aspects that work and those that, with a few recommendations, can soon work well too.

\subsection{Beyond Self-Promotion: We Already Do Some Things Right}

Sharing is caring. And we found that there is already a small, caring community that has drawn attention to at least 79 SE papers in LinkedIn posts for the two largest SE conferences alone. But not only that: the LinkedIn posts come from both paper authors and non-paper authors, they come from people in academia as well as people in industry. 

About 34\% of LinkedIn posts elaborate on or discuss the results of the respective studies.
This is great to see because such post intentions go beyond exclusively promoting oneself with one's own content on LinkedIn. While self-promotion is a legitimate concern, on its own, it barely serves science communication.

Nevertheless, congratulations on the acceptance of your paper! About 54\% of all comments on LinkedIn posts congratulated the poster. This happens especially when the poster has an academic background or is one of the paper authors. 
If practitioners commented (which was the case in 71\% of the comments), then the intentions of the comments were clearly more often of the nature that a discussion around the scientific content could arise. We consider both types of comments to be valuable: those in which people support and congratulate each other and those in which people exchange ideas and learn from each other. For the one half of the posts that received no comments at all, surely the creators would have been grateful to receive any.

Overall, we consider the current state of SE SciCom on LinkedIn to be better than we expected before conducting the study. In particular, the participation of practitioners, both regarding post creation and commenting activities on posts, is high.

\subsection{This Is How We Can Achieve Even More}
\begin{figure}[H]
    \centering
    \includegraphics[width=\columnwidth]{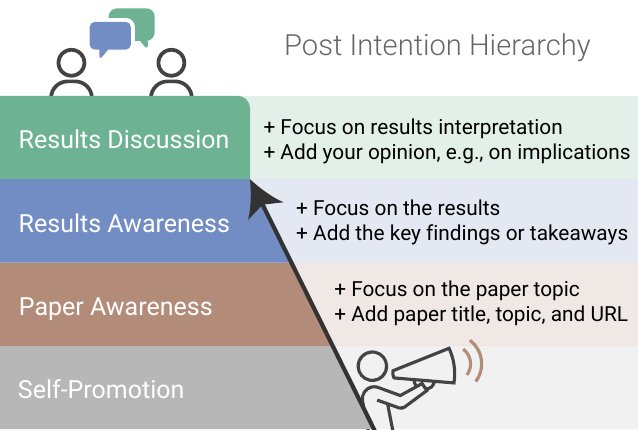}
    \caption{Synthesized post intention hierarchy for SE SciCom, with suggestions on how to move from one level to the next}
    \label{fig:post-intention-hierarchy}
\end{figure}

But even if some things are better than expected, what should SE researchers and practitioners do now to improve the state of the metaphorical SciCom bridge on LinkedIn?
How should they behave to make SE SciCom even more effective and mutually advantageous?
We have several suggestions, but believe that the synthesized post intention hierarchy plays an important role here (see Fig.~\ref{fig:post-intention-hierarchy}).
However, we first want to emphasize again that \textit{self-promotion} is not an inherently bad thing to do.
On the contrary, making a name for yourself and establishing yourself in a certain research field or topic is especially important for junior researchers who are thinking about an academic career.
Moreover, publicly celebrating successes is an important psychological countermeasure in an academic world that is far too often filled with rejections and negative feedback.
Lastly, the hierarchical nature of post intentions implies that even the higher levels like \textit{results discussion} still include \textit{self-promotion}, albeit to a very small degree.
With that being said, we think that it would be unfortunate if \textit{self-promotion} remains the only intention that a post conveys, due to the lost potential for effective SE SciCom.

We know that SE researchers are as over-worked as most academics, but we strongly believe that not much additional effort is required for basic SciCom.
Furthermore, many researchers already spent effort on sharing posts related to their papers on LinkedIn.
In these cases, it would be even less effort to slightly improve these posts for more effective SciCom.

Our first suggestion for researchers is therefore to try to \textbf{climb up the post intention hierarchy} when sharing SE research on LinkedIn.
As Fig.~\ref{fig:post-intention-hierarchy} describes, reaching \textit{paper awareness} is fairly low effort by adding the paper title, one to two sentences on the topic, plus a publicly available URL to the paper, e.g., to a preprint.
However, we encourage the SE community to at least aim for \textit{results awareness} by adding the key findings or takeaways, especially if you perceive them as industry-relevant.
Lastly, reaching \textit{results discussions} could be as simple as adding two to three more sentences that provide an interpretation of the results or that discuss the industry implications.
Remember to keep it brief, write in concrete language, and use headings to provide structure if necessary.
Sharing SE research in this way will make it much easier for practitioners to engage with the topic, or as \citet{Beecham:2014:MakingSERelevant} put it:
\enquote{researchers need to engage practitioners in a dialogue beyond that of observer and subject or author and reader. Blogs and discussion forums provide opportunities for researchers and practitioners to engage in informal discussions.}

One interesting finding was that we rarely saw any constructive interactions through comments between academics.
Our second suggestion for researchers is therefore to \textbf{not only congratulate, but engage with the topic} (if it is applicable for you).
Voice your support for its relevance, discuss its implications, ask questions about the study, or voice constructive criticism.
This not only improves interpersonal relationships in the community, but also contributes to more effective SE SciCom.
A nice example of this is provided by the post with \href{https://www.linkedin.com/posts/fabio-massacci-b3199225_a-benchmark-comparison-of-python-malware-activity-7008098845712654336-DxBn/}{\color{blue}ID 25}, where a more senior researcher reposts a shared ICSE paper of a junior researcher with his own positive interpretation, thereby adding value to the discussion and using his larger network for greater reach.

Finally, we do not have many suggestions for practitioners, as they already seem fairly engaged.
We only want to encourage them to keep commenting and discussing, and to seek exchanges on topics that are relevant to them.
This likely helps other practitioners, and can also be useful feedback for the researchers, maybe even lead to potential collaborations, particularly if the post author is a paper author.
So, while the SE SciCom bridge definitely needs support from both sides, we clearly see that academic SE researchers have a significant role in enhancing the strength of this connection.
Encouraging active engagement from both practitioners and researchers, we aspire to fortify the SE SciCom bridge, ensuring it becomes a beacon of excellence that advances the collective knowledge and impact of the SE community.

\section{Conclusion}
\label{sec:conclusion}

Throughout the paper, we spoke of a metaphorical bridge that is intended to bridge a gap between the scientific world and the practical world. A bridge where researchers and practitioners meet to discuss current research findings that can then be applied in the real world. Indeed, we have observed occasional cases on LinkedIn where researchers have communicated research findings and have been heard by their target audience. However, we would like to adjust the picture painted a bit to the actual situation reflected by our data. These data tell the story of a few practitioners crossing the bridge to later return to their practitioner communities with newly acquired research findings in their repertoire, where these findings are then discussed.

This is not what we expected, but the situation shows that part of SE research can be so interesting that practitioners take on the science communicator role. A role traditionally ascribed to researchers, but in the case of SE SciCom on LinkedIn also a role that is currently inadequately filled by researchers.

In this paper, we discussed how we can improve this situation and thus the exchange of SE research findings, based on a qualitative and quantitative study of 98 LinkedIn posts and their comments. 
Since the topic of science communication and the people involved are as exciting as they are complex, we call for further studies to be conducted.
A conversation with people who actively share SE research through social media platforms like LinkedIn would be a great way to find out more about their motivations and how they can be supported, especially if they do not have an academic background.
We also believe follow-up studies on predictor variables for the success of individual SciCom activities would help to demonstrate to more researchers that they themselves can determine the impact of their research to some degree.

\section*{Data Availability}

For transparency and reproducibility, we share our study artifacts online.\footnote{\url{https://doi.org/10.5281/zenodo.8410796}}
This includes tabular overviews of the included LinkedIn posts, their associated papers and all extracted attributes. Our source code for scraping and data analysis is included as well.

\bibliographystyle{ACM-Reference-Format}
\bibliography{main}

\end{document}